# Towards Nomadic 6G Communication Networks: Implications on Architecture, Standardization, and Regulatory Aspects

Daniel Lindenschmitt[1,†], Marcos Rates Crippa[2,†], and Hans D. Schotten[3,*†]

[†]Institute for Wireless Communication and Navigation. RPTU Kaiserslautern-Landau. D-67663 Kaiserslautern.

[1]daniel.lindenschmitt@rptu.de, [2]m.crippa@rptu.de, [3]schotten@rptu.de

## Abstract

The emergence of nomadic mobile communication networks for sixth-generation (6G) introduces a paradigm shift in how network infrastructure is conceptualized, deployed, and operated. Unlike traditional fixed systems, Nomadic Networks (NNs) consist of mobile and self-organizing nodes that provide radio infrastructure capabilities in motion. This paper explores the architectural implications of such systems, with a particular focus on the design and evolution of network interfaces. We analyze the requirements for inter-node communication, service discovery, and control delegation in dynamic environments. Furthermore, we examine the regulatory and licensing challenges that arise when infrastructure elements traverse jurisdictional boundaries. Based on current 6G visions and relevant research, we identify limitations in existing architectures and propose a set of interface principles tailored to nomadicity. By synthesizing findings from mobile, non-terrestrial, and organic network domains, this work contributes to the architectural foundation for future nomadic 6G communication systems and outlines directions for interface standardization in decentralized, mobile infrastructures.

## 1 Introduction

Nomadic 6G networks can constitute a paradigm shift in mobile communication architecture. So far, in mobile communication systems mobility was only possible for user devices and infrastructure remains static. NNs, especially in the domain of non-public networks (NPNs), introduced in the fifth-generation (5G), allow infrastructure elements such as access points, base stations, and even core components to move. This shift creates new dynamics in network topology, service availability, and connectivity assumptions. Traditionally, mobile networks are designed around the stability of Core Network (CN) and Radio Access Network (RAN) components. Network planning, interface definition, and protocol operation rely on long-term availability of static links and predictable coverage areas. The ability to endure temporary loss of backhaul connection enables infrastructure that can frequently change location and also provides benefits of new types of backhauling such as nonterrestrial connections.

To enable this technically, interfaces such as N2 and N3 must evolve beyond their original assumptions. This includes dual-mode interface operation, allowing localized execution of core functions like session and mobility management when backhaul is unavailable. Integration with container-based platforms and orchestration tools (e.g., HashiCorp Nomad) supports dynamic deployment of network functions within mobile nodes, enabling autonomous operation until reconnection. These requirements particularly affect network interfaces, which are traditionally built on high stability, continuous connectivity, and clear role separation. Interfaces like X2, N2, N3, and other NG-C/NG-U links were not designed to handle mobile infrastructure, dynamic topology discovery, or peer coordination. The novelty of this work lies not only in consolidating these challenges but in proposing specific adaptations to current interface structures and deployment frameworks. In addition to technical challenges, NNs raise complex questions around regulation and licensing. For example, there is no harmonized framework for NPN spectrum allocation. Existing regulatory models are based on geographically fixed spectrum, which conflicts with the mobility of nomadic infrastructure. Regulatory frameworks must evolve to accommodate this by adopting dynamic licensing, spectrum sharing, and inter-domain coordination. Concepts such as spectrum entitlement tokens, localized license servers, and inter-domain roaming are discussed in this work as possible enablers of lawful nomadic operation. Starting with Section 2, we introduce recent research in the field of regulation and architectural approaches towards 6G. Section 3 explores challenges while realizing nomadic 6G networks and outlines directions for future network interface design. Section 4 discusses required adaptations in 6G standardization and regulation, followed by Section 5 with a conclusion and outlook.

## 2 Related Work

Research on mobile infrastructure and NN architectures has gained increasing attention in the context of 6G and beyond. The concept of infrastructure mobility was introduced in early works on vehicular and drone-assisted networks, where base stations are mounted on mobile platforms to extend coverage and capacity [1-5]. These systems demonstrated the feasibility of moving network elements with focus on the 5G gNodeB Basestation (gNB), but often assumed persistent backhaul and centralized control. The concept presented in the following is a possible way of making the entire cellular network, both RAN and CN, ready for a mobile usage.

The architectural foundations of NNs have recently been explored in the context of 6G. A comprehensive analysis of architectural challenges for nomadic infrastructure is provided in [6], identifying critical gaps in connectivity assumptions, function placement, and mobility support. Complementary to this, the use of nomadic non-public networks (NNPNs) has been proposed for industrial and tactical applications, highlighting use cases and key performance indicators that differ significantly from traditional public networks [7]. New fields of application have been identified, which can only be made possible by NNPNs. NNPNs hereby represent a subgroup of the more general NNs. Further architectural perspectives are discussed in the vision of organic 6G networks [8], which proposes selforganizing and self-adaptive systems capable of handling the dynamic behavior of nomadic nodes. These concepts align with the need for decentralized and resilient infrastructure, especially in non-terrestrial and mobile environments. The emerging paradigm of 3D and unified terrestrial-nonterrestrial networks also intersects with nomadic concepts. Recent work presents a unified 3D network architecture that incorporates ground, aerial, and satellite components into a coherent mobile communication fabric [9]. The mobility of network elements across all three spatial dimensions introduces similar challenges to those in nomadic systems, particularly in terms of interface design and interference management. From an interface perspective, prior studies have examined requirements for dynamic and flexible inter-node communication. Delay-tolerant and opportunistic networks provide mechanisms for asynchronous message passing and federated control [10, 11], which are relevant when infrastructure elements exhibit intermittent connectivity.

Efforts toward federation and decentralization are further exemplified in [12], where distributed 5G core functions are explored to enable mobile or temporary node clusters. The interface implications of such systems include dynamic role negotiation and on-demand service discovery, both of which are central to nomadic infrastructure. The broader context of 6G also envisions capabilities beyond traditional communication. Research highlights the integration of sensing, localization, and compute services, which demand significant flexibility in architectural design [13]. These multipurpose roles reinforce the need for adaptable interface specifications that can dynamically support heterogeneous functions.

Regulatory and licensing challenges are also gaining attention. Classical spectrum allocation and jurisdictional models are inadequate for nomadic deployments, especially when nodes traverse administrative boundaries or utilize dynamic spectrum access [14, 15]. Solutions such as regulatory sandboxes and dynamic spectrum authorization have been proposed [16] but require further standardization to ensure interoperability. In tandem with the accessibility of affordable hardware and uncomplicated operation, a simple and flexible approval process is a decisive factor. However, the current legal framework is not yet equipped to deal with nomadic scenarios [17].

This body of research establishes the foundation for nomadic mobile communication networks and highlights a growing interest in decentralized and mobile architectures. However, there remains a significant gap in the design of standardized interfaces that are capable of supporting nomadicity. This work addresses this gap by proposing architectural principles focused on network interfaces for mobile infrastructure.

## 3 Realization Challenges of Nomadic Networks

The NPNs introduced by the current 5G standard operate as independent stationary private networks, in contrast to the NNPNs proposed here, which have mobility and flexibility as defining characteristics. As a consequence of these new features, NNPNs will face challenges not yet addressed by the 6G standardization in the scope of architectural requirements and spectrum regulations established for 5G. In this section both aspects are investigated. Regarding architectural challenges an investigation of relevant network interfaces will take place. Afterwards, we analyze how spectrum is regulated in general and for NPNs specifically and examine the challenges NNPNs will introduce in this area.

### 3.1 Architectural Implications

#### 3.1.1 Network Interfaces

The concept of NNs introduces fundamental architectural challenges for 6G, particularly in relation to the design and function of network interfaces. Unlike traditional cellular deployments, where infrastructure is fixed and consistently connected, NNs require network elements to remain functional despite frequent changes in location, intermittent connectivity, and the absence of continuous backhaul. These characteristics impose specific requirements on the N2 and N3 interfaces, the two primary links between the gNB and the CN. The N2 interface connects the gNB to the Access and Mobility Management Function (AMF) and facilitates control plane signaling, including registration, session management, and mobility procedures. In contrast, the N3 interface connects the gNB to the User Plane Function (UPF) and carries user traffic. Both interfaces were originally designed under the assumption of stable, low-latency backhaul and continuous connectivity to the core. As such, they are not sufficiently robust for nomadic scenarios. When infrastructure nodes operate in isolation or as dynamic clusters, new strategies for both control-plane and user-plane interface behavior become necessary.

A key limitation of current architecture is its reliance on centralized control. In typical deployments, gNBs offload all session and mobility management to the AMF over the N2 interface. This assumption fails in a nomadic setting. A gNB deployed on a vehicle, drone, or vessel may lose access to the central AMF for extended periods, rendering centralized mechanisms inoperable. To address this, nomadic nodes must be equipped with localized control capabilities. Specifically, gNBs should host lightweight AMF-like functionality to autonomously manage session state, mobility, and access control, as illustrated in Figure 1. These localized functions must operate independently during disconnection and synchronize with the central AMF upon reconnection. The N2 interface must therefore support dual-mode operation: facilitating centralized signaling when connected, while enabling local decision-making during isolation. Furthermore, mechanisms are needed to reconcile local and global state. Actions such as handovers, session updates, or policy changes executed during disconnection must later be integrated into the CN's state. To support this, the N2 interface should include lightweight synchronization and reconciliation protocols that re-merge the state with minimal overhead.

The N3 interface faces similar challenges in the user plane. In nomadic deployments, continuous user-plane connectivity to the CN cannot be guaranteed. When the N3 link is unavailable, the gNB must implement local data handling, including buffering, delay-tolerant forwarding, or temporary proxy-UPF functionality, as shown in Figure 1. Such enhancements require architectural modifications to the gNB and the N3 interface so it can adapt to varying link quality and availability. During full disconnection, the gNB may store user traffic locally until reconnection. During partial disconnection or degraded backhaul, traffic may be relayed through neighboring nomadic nodes that retain access to the core. This calls for a multi-hop user-plane architecture in which N3-like functionality is extended across dynamically formed mobile clusters. Unlike fixed deployments—where paths from gNB to UPF are predetermined—NNs must dynamically discover and manage data paths. To enable this, the N3 interface must evolve to include support for route discovery, path management, and retransmission strategies that are resilient to mobility and topology changes. Conventional forwarding mechanisms based on static routes must be replaced by context-aware, mobility-tolerant frameworks.

### 3.1.2 Physical Deployment Options

Another key consideration is the separation of interface logic from physical transport. A NN can operate as a fully autonomous system, hosting its own gNB, RAN, and CN. This approach offers maximum independence but increases hardware demands. More commonly, hybrid deployments are used—where parts of the RAN are mobile, while other functions remain remote or centralized [6]. This requires robust communication between distributed components. In nomadic environments, the backhauling link can vary from terrestrial to non-terrestrial satellite links, each with different latency, jitter, and reliability profiles. As a result, interface logic must be abstracted from the transport layer. The N2 and N3 interfaces should function as logical con-structs operating atop adaptable transport layers. They must dynamically select and switch between available physical bearers, based on link conditions. This virtualization of transport requires built-in monitoring and adaptation mechanisms. For example, the system should detect link degradation, adjust interface behavior, and notify peer entities of connectivity changes. Interfaces must adapt protocol parameters—such as retransmission timers, flow control windows, and acknowledgment schemes—based on real-time transport conditions.

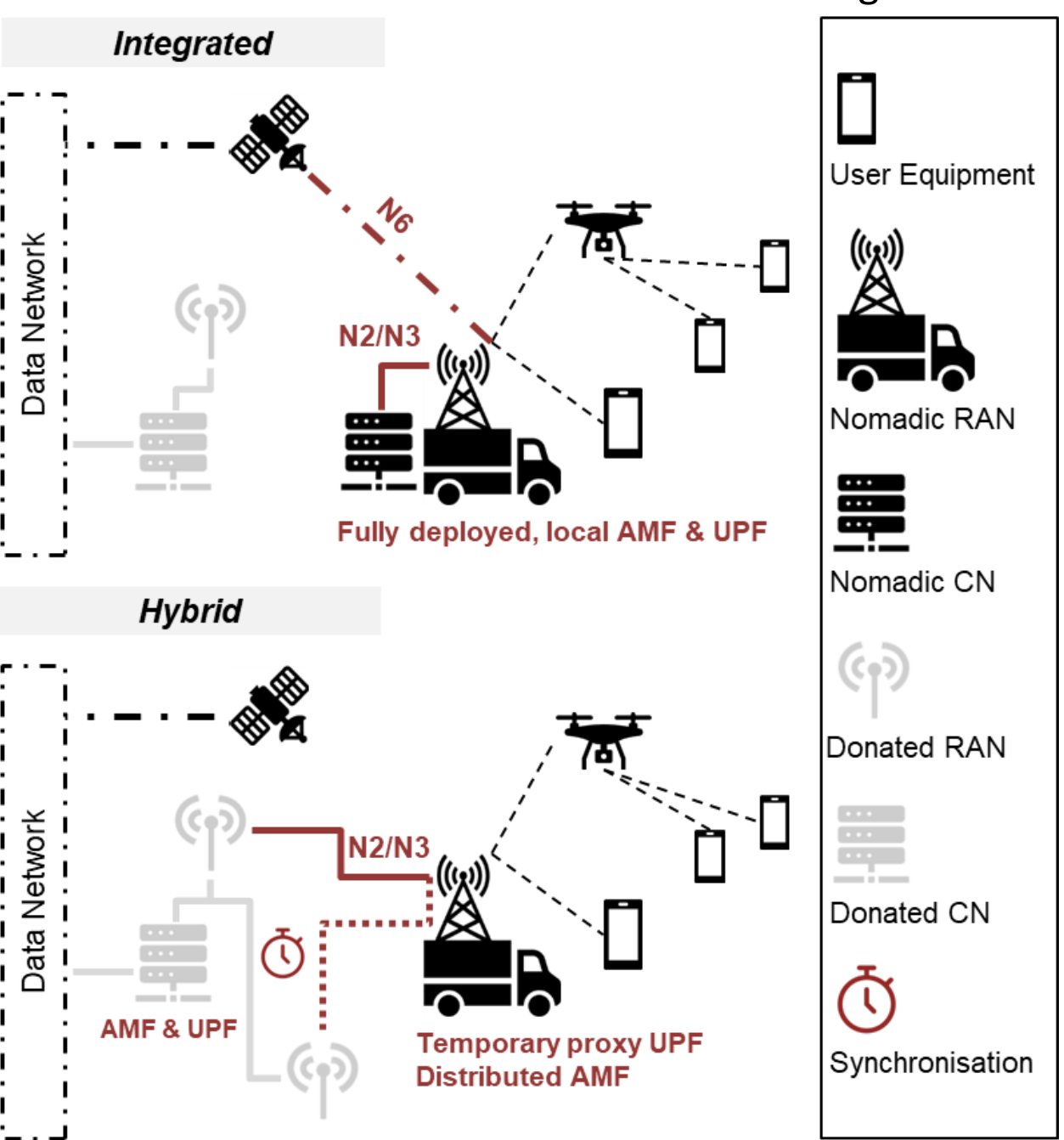

**Figure 1 Integrated and Hybrid Mode of Nomadic Networks**

## 3.2 Aspects of Regulation and Licensing

### 3.2.1 General Overview

Spectrum regulation entails the organization, control, and monitoring of the utilization and allocation of frequency bands. The radio spectrum is a natural resource. In order to optimally utilize the spectrum, coordination and management are required to ensure efficient use while avoiding interference between radio communication services. 3rd Generation Partnership Project (3GPP) standards and International Telecommunication Union (ITU) guidelines, which define the bands that will be available for each radiocapable technology, are created to fulfill those requirements. Supranational organizations (such as the EU) and national governments use these guidelines as the basis for their local regulations. Regulators aim for the optimal utilization of spectrum for the social, cultural, and economic benefit of society. The EU uses the European Electronic Communications Code (EECC) to provide the legal framework for its member nations. Within and outside of the EU, each individual country has the final say regarding the licensing and monitoring of the spectrum accessible within their borders, just as with other natural resources [18]. National regulators have three methods for assigning spectrum [19]: *Administrative Allocation*: This refers to the fixed allocation of spectrum by a regulator, based on its own criteria. Interested parties can either be directly awarded or be the winner of closed governmental bidding processes, which is the most traditional method, and has been used for 3G and 4G spectrum allocation. In 5G, administrative allocation is currently used for local networks (e.g. in Germany) and occasionally for awarding spectrum to MNOs (e.g. in Japan) [20]. Administrative allocation has been criticized for its lack of flexibility, inefficiency, slow response to technological and market changes, and the unfair role that political influence can play in the regulator's decision making. It will most likely play a lesser role in 6G, especially when compared with the other two methods mentioned next. *Market-based Mechanisms*: This method consists of national regulators letting the market make decisions about spectrum allocation, usually through hosting spectrum auctions to award spectrum property rights. Any interested party can bid for spectrum, and the winners for a certain frequency band will get a license to operate in that frequency within an area (usually nationwide) for a certain period of time. Spectrum owners must follow regulations and conditions established in their license agreements, and are allowed to trade and lease their acquired spectrum to third parties. The primary motivator for holding spectrum auctions is to improve spectrum usage, since it is assumed that any party willing to pay for a frequency band will work to extract as much value as possible, to justify the initial buying cost. Another advantage is that the revenue governments can accrue with the selling of spectrum. Auctions have been used for allocating spectrum for 3G, 4G and 5G frequency bands. It is expected for them to remain a preferred method for governments, not only due to aforementioned advantages, but also due to the more complex and wider range of bands that 6G is expected to operate on. The flexibility market-based mechanisms exhibit is seen as better suited for the scenario. The main criticism levied against spectrum auctions is that bidders can manipulate them either through colluding among themselves or overbidding to guarantee they will win. Future offerings of spectrum will have to take into account these market manipulations and make sure the whole system is fairer. *Unlicensed Commons*: Since the number of radio-capable devices and technologies can be overwhelming to regulators, one strategy is to not require any license for a selected group of frequency bands. Radio devices have to follow rules and regulations to operate in those frequency bands, but there is no exclusive access to them. This means devices must coordinate among themselves. Wi-Fi services are the most notable example, using the unlicensed 2.4 GHz, 5 GHz and 60 GHz bands. The main advantage of this approach is the lack of any significant barrier to entry. Services do not need to go through any lengthy licensing process, they can just start using the spectrum as soon as possible. Easy access to the unlicensed spectrum can lead to a crowded space with many radio devices, leading to the possibility of interference. Mechanisms such as Listen-Before-Talk (LBT) can be used to ensure a fair and harmonious coexistence, but ultimately the whole scenario might be too dynamic for services that require strict and high QoS requirements. It is possible for cellular technologies to operate within unlicensed spectrum, using technologies such as Long Term Evolution Unlicensed for 4G and New Radio Unlicensed for 5G.

**Table 1 Overview of spectrum assignment and allocation for the 3.5 GHz band**

| Country | Method of Allocation | Description |
|---|---|---|
| Finland | Market-based Approach | Spectrum awarded through auctions only to existing Mobile Network Operators (MNOs). Local networks have to be established by a MNO or lease the spectrum from them. |
| Germany | Administrative Allocation | Applications must be made directly to the government for local licenses. |
| Japan | Administrative Allocation | Licenses for local private 5G indoor networks are issued by the government for a fee. MNOs are not allowed to apply for local 5G licenses. |
| UK | Administrative Allocation | Licensing only for low and medium power base stations under a Shared Access process, on a first-come first-served basis. |
| USA | Unlicensed Commons | The lowest tier of CBRS (referred to as General Authorized Access) allows open access to the band. Devices must avoid interfering with higher-tiered incumbents. |

#### 3.2.2 Current Regulations for NPNs

Multiple countries have allocated spectrum for NPNs, though no global harmonization exists. Frequently used bands include n258 and n40, with some convergence across regions—e.g., many European countries assign spectrum in the 26–28 GHz range [20]. A comparative analysis of local 5G regulation across five countries (Finland, Germany, Japan, UK, and the USA) for the 3.5 GHz band is provided in [20]. Table 1 summarizes the key findings.

#### 3.2.3 Primary Regulatory Challenges for Nomadic Networks

There is currently no fast and direct way to acquire exclusive frequency bands. This represents the main challenge facing NNPNs when it comes to spectrum. Even if spectrum is available in a certain region, any stakeholder wishing to deploy a NNPN will need to either go through an application request with the government or a leasing agreement with a MNO. Both of those processes are currently conducted manually (meetings, emails, phone calls) [21]. The whole process might take weeks or even months. To address this, a system for digitally pre-authorized spectrum entitlements can be introduced. A nomadic node could carry a signed token proving conditional spectrum access, issued by a national or supranational body. Upon entering a new region, the node would query a local policy server or coordination system (similar to CBRS SAS or LSA controllers) to validate entitlement and request activation. This mechanism supports dynamic operation and reduces human overhead. Another major challenge is cross-border operation. Neighboring countries may have vastly different regulatory frameworks, and frequency assignments may not align. To support international mobility, a federated licensing framework is needed—one where infrastructure operators register their nodes and receive recognition across jurisdictions via roaming-like agreements. On the network side, nomadic nodes must be equipped with hardware and software that allow flexible reconfiguration based on allowed frequencies and transmission parameters. If no suitable band is available in a given location, the network may not be deployable, undermining reliability and coverage. These regulatory barriers directly influence interface design and system architecture. Spectrum access mechanisms must be tightly integrated with control-plane signaling to allow real-time compliance updates. For example, when a node changes geographic area, it must update its spectrum configuration and inform peer nodes and control functions of the change.

While NPNs have shown how localized networks can be regulated within current systems, NNPNs require a fundamentally more dynamic approach. Practical mechanisms such as entitlement tokens, inter-domain authorization, and dynamic spectrum APIs offer feasible solutions. These will need to be standardized in future 6G releases and supported through international regulatory cooperation.

# 4 Evaluation

## 4.1 Impacts of NN Requirements on 6G Standardization

The architecture of 6G must evolve significantly to support nomadic mobile networks. Central to this transformation is the adaptation of the N2 and N3 interfaces. These interfaces, originally designed for fixed deployments in 5G, must be extended to support decentralized and context-aware communication between mobile infrastructure components. Current 5G architecture assumes that gNBs interact mainly with the core. In contrast, NNs require peer-to-peer operation. Infrastructure nodes must directly coordinate in the absence of centralized management. To this end, the N2 interface must support signaling between peer gNBs for handover orchestration, session transfer, and dynamic role negotiation. A peer-N2 extension requires mechanisms for dynamic node discovery, mutual authentication, and secure signaling between distributed mobile nodes. These capabilities build on, but go beyond, existing Xn interfaces in 5G. The N3 interface must similarly evolve to support data relaying between nodes. For example, if a gNB loses access to its UPF, it should be able to route user traffic via a neighboring node. In addition, temporary proxy-UPF functions must be available to anchor user sessions locally, forming a distributed user-plane fabric. These requirements are aligned with the broader trend toward function virtualization and service-based architecture (Service-Based Architecture (SBA)). Core functions, including AMF and UPF, must be deployable as containerized microservices on edge or mobile infrastructure, with interfaces that adapt to changing topologies and backhaul quality. Security and trust are fundamental in this model. Peer coordination increases the attack surface, requiring interfaces to incorporate lightweight encryption, local key management fallback, and node authentication. Distributed Ledger Technology (DLT) can support these requirements by offering decentralized "Trust-as-a-Service" functionality [22, 23]. Table 2 summarizes how each requirement impacts interface design and highlights proposed extensions and implementation strategies:

**Table 2 Interface Adaptations for Nomadic Network Requirements**

| Requirement | Affected Interface | Proposed Extension |
| --- | --- | --- |
| Local control during disconnection | N2 | Embedded AMF, dual-mode N2, local fallback logic |
| Peer-to-peer coordination | N2 | Peer-N2 support, secure signaling, node discovery |
| User-plane relaying | N3 | Multi-hop forwarding, proxy-UPF, session anchoring |
| Adaptive con-nectivity | N2/N3 | Virtual interface abstraction, transport-aware signaling |
| Security and trust | N2/N3 | Local authentication, lightweight encryption, DLT anchors |

## 4.2 Proposed Approaches for Spectrum Allocation for NNs

Based on current spectrum allocation models, three principal strategies can be applied to nomadic deployments:

*Licensing via MNOs*: Stakeholders may lease spectrum from mobile operators under agreed conditions. The lease may be exclusive or follow shared access principles (e.g., LSA) [23].

*Direct licensing by regulators*: Similar to local 5G licensing, spectrum could be granted directly to nomadic operators with temporal and geographical constraints. *Unlicensed operation*: The most flexible approach, enabling instant deployment without centralized negotiation, but with minimal QoS guarantees.

Each method has trade-offs. Unlicensed access is fast but unreliable; licensed access provides performance guarantees but requires pre-planning. A middle ground is dynamic spectrum access: pre-authorized operators submit real-time usage requests to a centralized or distributed monitoring system, which grants short-term licenses based on availability. Such a system could expose standardized spectrum APIs for automated query, request, and authorization, enabling low-latency spectrum allocation suitable for mobile and dynamic deployments. Regulatory bodies could enforce constraints (e.g., power, duration, location) via programmable policies. This mechanism ties into the architecture discussed earlier. A nomadic node that detects location change could trigger a spectrum request over this API, update its internal configuration, and notify peer and core components via control-plane signaling.

# 5 Conclusion

This paper analyzed the architectural and regulatory implications of nomadic mobile communication networks within the context of future 6G systems. Limitations of current 5G architectures, particularly regarding the N2 and N3 interfaces—were identified, and necessary enhancements were outlined to support infrastructure mobility. To enable reliable operations in dynamic environments, proposals included dual-mode operation for N2, allowing localized control during core disconnection, and extensions to N3 for user-plane relaying, multi-hop forwarding, and session anchoring in mobile deployments. The analysis further addressed spectrum regulation challenges, emphasizing the incompatibility of static licensing frameworks with nomadic operation. A dynamic, location-aware spectrum entitlement model was proposed to facilitate lawful, efficient use of frequency bands across jurisdictions, based on real-time policy enforcement and entitlement token validation. A structured evaluation highlighted how these adaptations align with key functional requirements, including local autonomy, peer-topeer coordination, and secure decentralized operation. Integration of orchestration frameworks and spectrum APIs into the control plane was identified as a critical enabler for real-time adaptation in nomadic scenarios.

Future work should investigate methods to improve scalability and security, particularly for large-scale deployments. Integration with edge computing and emerging 6G technologies can expand the capabilities of NNs. In addition, regulatory frameworks must evolve to accommodate mobile infrastructure and dynamic spectrum use. Field studies in diverse environments will be essential to validate the performance, usability, and practicality of the NNs approach. These efforts will support the development of nextgeneration mobile networks that are more adaptable, resilient, and user-driven.

# Acknowledgment

The authors acknowledge the financial support by the German *Federal Ministry for Research, Technology and Space (BMFTR)* within the project »Open6GHub« {16KISK003K & 16KISK004}.